\documentclass[%
 reprint,
 amsmath,
 amssymb,
 aps,
prl,
 twocolumn
]{revtex4-1}

\usepackage{comment}

\usepackage{cancel}
\usepackage[normalem]{ulem}

\usepackage{lipsum, babel}
\usepackage{graphicx}
\usepackage{dcolumn}
\usepackage{bm}
\usepackage{appendix}
\usepackage{comment}
\usepackage{cancel}
\usepackage{graphicx}
\usepackage{dcolumn}
\usepackage{bm}
\usepackage{algorithm}
\usepackage{algpseudocode}
\usepackage{capt-of}
\usepackage{xcolor} 

\usepackage[normalem]{ulem} 

\usepackage[normalem]{ulem} 
\usepackage{amsmath}
\usepackage{tikz-cd}
\usepackage{xr}
\begin{document}


\renewcommand{\ULthickness}{2.pt}

\newcommand{\kBT}{{k_\mathrm{B} T}}
\newcommand{\vecF}{\boldsymbol{f}_\mathrm{nc}}
\newcommand{\veczeta}{\boldxi}

\newcommand{\st}{{\mathrm{st}}}
\newcommand{\rhost}{\rho^\mathrm{st}} 
\newcommand{\boldvst}{\boldsymbol{v}^\mathrm{st}} 
\newcommand{\vst}{{v}^\mathrm{st}} 
\newcommand{\boldFzero}{\boldsymbol{F}_0}

\newcommand{\boldA}{\boldsymbol{A}}

\newcommand{\boldr}{\mathbf{r}}

\newcommand{\Lgamma}{\mathcal{L}_\gamma}

\newcommand{\boldeta}{\eta} 
\newcommand{\boldv}{{\boldsymbol{v}}} 
\newcommand{\boldlambda}{{\boldsymbol{\lambda}}}
\newcommand{\boldx}{{\boldsymbol{x}}}
\newcommand{\trajx}{\boldsymbol{x}}
\newcommand{\boldc}{c}
\newcommand{\dt}{{\,\mathrm{d}t}}
\newcommand{\dvarphi}{{\,\mathrm{d}\theta}}
\newcommand{\ds}{{\,\mathrm{d}s}}
\newcommand{\dx}{{\,\mathrm{d}\boldsymbol{x}}}
\newcommand{\dee}{{\mathrm{d}}}
\newcommand{\gammadot}{{\dot{\gamma}}}
\newcommand{\boldnabla}{{\boldsymbol{\nabla}}}
\newcommand{\gammaperp}{{\Sigma}}

\newcommand{\boldFnc}{{\boldsymbol{F}_\mathrm{nc}}}
\newcommand{\Fnc}{{{F}_\mathrm{nc}}}
\newcommand{\Fncsquared}{{{F}_\mathrm{nc}^2}}
\newcommand{\Fncinverse}{{{F}_\mathrm{nc}^{-1}}}

\newcommand{\boldxi}{{\boldsymbol{\xi}}}
\newcommand{\boldzero}{{\boldsymbol{0}}}

\usetikzlibrary{cd}
\makeatletter
\tikzcdset{
  eq node/.style={
    commutative diagrams/math mode=false, anchor=center},
  eq/.style={
    phantom,
    /tikz/every to/.append style={
      edge node={node[commutative diagrams/eq node]
        {\@eqnswtrue\make@display@tag\ltx@label{#1}}}}}}
\makeatother

\title{Optimal active engines obey the thermodynamic Lorentz force law}
\author{Adrianne Zhong$^{1,2,3}$}
\email{adrianne.zhong@northwestern.edu}
\author{Adam G. Frim$^{2,4}$}
\author{Michael R. DeWeese$^{2,3,5}$}
\affiliation{%
 ${^1}$NSF-Simons National Institute for Theory and Mathematics in Biology, Chicago, IL, 60611\\
 ${^2}$Department of Physics, University of California, Berkeley, Berkeley, CA, 94720 \\
${^3}$Redwood Center For Theoretical Neuroscience, University of California, Berkeley, Berkeley, CA, 94720 \\
  ${^4}$Department of Physics and Astronomy and the Center for Soft and Living Matter, University of Pennsylvania, Philadelphia, PA, 19104 \\
${^5}$Department of Neuroscience, University of California, Berkeley, Berkeley, CA, 94720
}%
\date{\today}

\begin{abstract}
What are the fundamental limitations for finite-time engines that extract work from active nonequilibrium systems, and what are the optimal protocols that approach them? We show that the finite-time work extraction for nonconservative overdamped Langevin systems may be rewritten as a Lorentz force Lagrangian action, with the kinetic term corresponding to a thermodynamic metric term that is an $L_2$-optimal transport cost for the time-dependent probability density, and the magnetic field coupling term corresponding to an effective quasistatic work extraction, proving that optimal protocols counterdiabatically steer the thermodynamic state trajectory to satisfy a Lorentz force law defined on thermodynamic state space. We utilize and reinterpret classic concepts from electromagnetism in the setting of cyclical nonequilibrium processes. We show that the housekeeping heat can be controlled to be arbitrarily close to zero by minimizing nonequilibrium fluctuations. It immediately follows from our results that the constant-velocity angle clamp protocol applied to the $F_1$ molecular motor in a recent experiment [T. Mishima \emph{et al.}, Phys. Rev. Lett. 135, 148402 (2025)] is in fact the globally optimal protocol: it produces zero housekeeping heat{, while simultaneously minimizing dissipation and maximizing work transduction.} 
\end{abstract}

\maketitle

\emph{Introduction.}---A natural hypothesis for biological systems is that, while fundamentally constrained by the second law of thermodynamics, they have evolved to be thermodynamically efficient in metabolizing energy \cite{schrodinger1992life}. However, fully characterizing efficient biological processes is challenging: unlike traditional, well-studied heat engine models dating back to the 19th century~\cite{carnot1824reflections}, biological systems operate far from equilibrium and actively break time-reversal symmetry ~\cite{england2013statistical, fodor2016far}. 

Stochastic thermodynamics provides a framework to study fluctuating nonequilibrium systems \cite{seifert2012stochastic}. Previous studies have demonstrated that within this framework, the dissipation of a driven passive system is purely geometric, revealing that optimal processes are geodesics in thermodynamic state space \cite{nakazato2021geometrical, zhong2024beyond}. (Initially, this result was shown as approximate in linear-response \cite{sivak2012thermodynamic}, but has been more recently generalized to be exact in terms of optimal transport geometry \cite{aurell2011optimal, ito2024geometric, nakazato2021geometrical}.)

In this Letter, we tackle the problem of cyclically extracting maximal work from active, nonconservatively-driven overdamped systems such as ATP-synthase \cite{junge2015atp}. Notably, we demonstrate a surprising fundamental relationship to the Lorentz force law of electromagnetism. 

The problem statement is: Given a configuration space $x \in \mathcal{X} \subseteq \mathbb{R}^d$, a fixed temperature $\kBT$, a fixed time-reversal-symmetry-breaking nonconservative vector field $\boldFnc(\boldsymbol{x})$ {(i.e., there does \emph{not} exist a $\phi(\boldx)$ satisfying $\boldFnc = -\boldnabla \phi$)}, a potential energy function $U_\lambda(\boldsymbol{x})$ parameterized by $\lambda \in \mathcal{M}$, and a fixed cycle duration $\tau$, what is the optimal $\tau$-periodic protocol $\lambda^*(t)$ that maximizes the cyclic output work {\cite{fodor2022irreversibility}}
\begin{equation}
  W_\mathrm{out}[\lambda(t)] =    -\oint_0^\tau \dot{\lambda}^i \bigg\langle \frac{\partial U_{\lambda}}{\partial \lambda^i } \bigg\rangle 
  \dt  \label{eq:work-out-definition} \ ?
\end{equation}
Here $\mathcal{M}$ is an $m$-dimensional manifold, and we use Roman indices to denote local coordinates on $\mathcal{M}$ and adopt Einstein summation notation for Roman indices. The ensemble average $\langle \cdot \rangle$ is a $\tau$-periodic nonequilibrium steady-state average \cite{branicki2021time} over stochastic trajectories $\boldx(t)$ obeying the overdamped Langevin equation
\begin{equation}
  \dot{\boldsymbol{x}} = -\mu \boldnabla U_{\lambda(t)}(\boldx) +  \mu \boldFnc(\boldx) + \sqrt{2 \mu \kBT} \, \boldxi(t), \label{eq:Langevin}
\end{equation}
where $\boldxi(t)$ is Gaussian white noise with unit amplitude, and $\mu$ is a constant motility.

{In contrast to previous studies of optimal protocols in active nonequilibrium systems that have focused on minimizing entropy production~\cite{zulkowski2013optimal, mandal2016analysis, davis2024active, soriani2025control, lacerda2025information}, here we consider maximizing the extracted output work (in physical units of energy) through cyclically varying $\lambda(t)$ at constant $\kBT$ and fixed $\boldFnc(\boldx)$. 
While previous studies have considered multiple heat sources at variable temperatures \cite{bo2013optimal, kumari2020stochastic, martin2018extracting, gronchi2021optimization, frim2022geometric, miangolarra2022geometry} or time-varying activity \cite{casert2024learning}, 
here we derive fundamental results that are valid arbitrarily far from equilibrium, for this simpler, unadorned setting \cite{isothermal}.}

A challenging aspect of this problem is that the nonequilibrium thermodynamic state of the system $\rho_t$ depends on the history of the protocol $\lambda(t')|_{t' \leq t}$. Prior work ({\it e.g.}, \cite{zulkowski2013optimal, gupta2023efficient, davis2024active}) has employed linear response theory to show that integrands like Eq.~\eqref{eq:work-out-definition} may be approximated by a geometric expression, \textit{e.g.,} for excess work $\dot{W}_\mathrm{ex} \approx \dot{\lambda}^i \dot{\lambda}^j g^\mathrm{lr}_{ij}(\lambda) + ..., $  involving a linear-response thermodynamic metric tensor $g^\mathrm{lr}_{ij}(\lambda)$ for $\lambda \in \mathcal{M}$. Formally, this is done through approximating the thermodynamic state $\rho_t \approx \rhost_{\lambda(t)} + \dot{\lambda}^i  h_i(\lambda(t))$ using an appropriate response function $h_i(\lambda)$ \cite{dambrosia2025higher} and is a valid approximation in the slow-driving regime (where $|\dot{\lambda}|$ is ``appropriately small'' \cite{wadia2022solution, sawchuk2024dynamical, dambrosia2025higher}); however, this approximation need not hold outside of this regime.

In this Letter, we set out to tackle the full nonequilibrium problem without resorting to any form of linear-response approximation. Through shifting the focus from performing linear response around the control parameter $\lambda(t)$ to considering the actual nonequilibrium thermodynamic state of the system $\rho_t$ itself, we demonstrate the \emph{exact} relation
\begin{equation}
    -W_\mathrm{out} = \oint_0^\tau \bigg[ \mu^{-1} \,\dot{\gamma}^i \dot{\gamma}^j g_{ij}(\gamma(t))\,  + \dot{\gamma}^i A_i(\gamma(t))  \bigg] \mathrm{d}t, \label{eq:w_out_result} 
\end{equation}
where $\gamma(t)$ parametrizes the $\tau$-periodic nonequilibrium thermodynamic steady state of the system, and $g_{ij}(\gamma)$ and $A_i(\gamma)$ are, respectively, a metric tensor and a differential one-form defined on thermodynamic state space. The thermodynamic metric $g_{ij}$ encodes the optimal transport geometry of the underlying probability density \cite{zhong2024beyond}, while the thermodynamic one-form $A_i$ can be expressed as an effective quasistatic conjugate force.

\begin{figure}
  \centering
    \includegraphics[width=.8
    \linewidth]{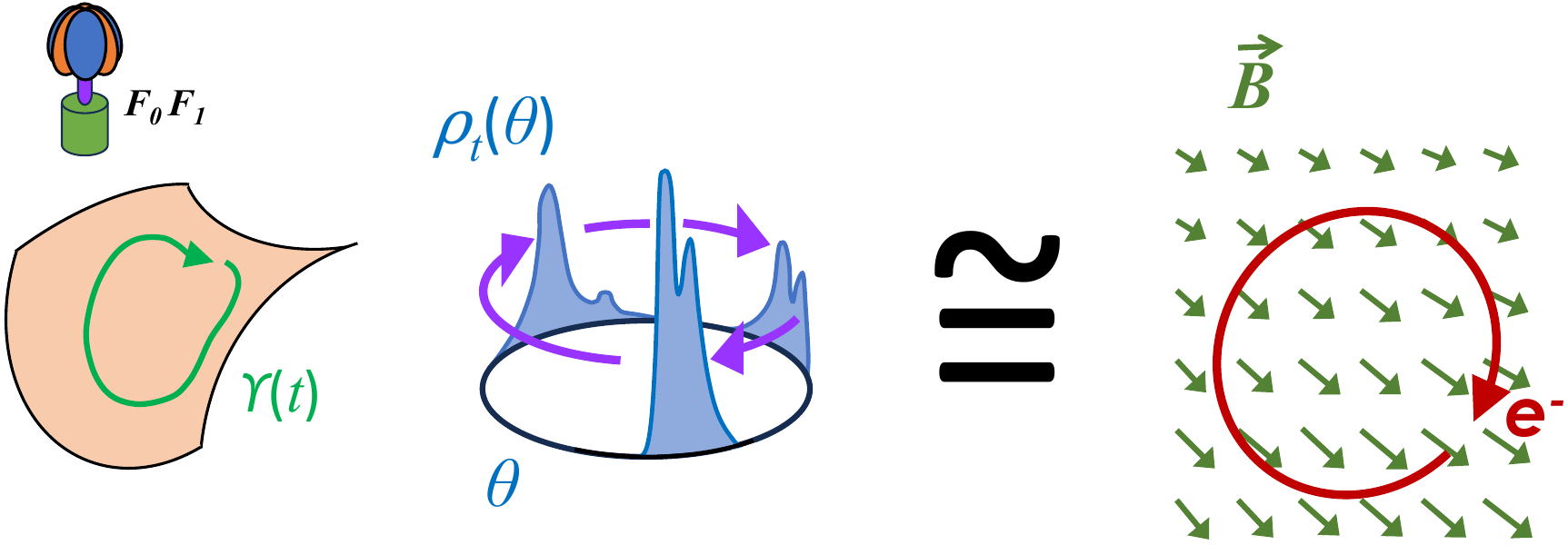}
    \caption{A cartoon depiction of our main result:~for~a cyclically controlled active system (such as $F_O F_1$ ATP-synthase), the output work is a Lorentz force Lagrangian integral [Eq.~\eqref{eq:w_out_result}] for a curve $\gamma(t)|_{t \in [0, \tau]}$ that parameterizes the time-dependent nonequilibrium thermodynamic state $\rho_t$; optimal thermodynamic state trajectories must satisfy a corresponding Lorentz force law, akin to trajectories of a charged particle in a magnetic field.}
    \label{fig1}
\end{figure}

Remarkably, the integrand in Eq.~\eqref{eq:w_out_result} takes the form of a Lorentz force law Lagrangian from classical electromagnetism (EM), and so \emph{globally optimal} thermodynamic state trajectories $\rho_t^*$ obey a Lorentz force law on thermodynamic space (Fig.~\ref{fig1}), and \emph{globally optimal} protocols are ones that realize these state trajectories via counterdiabatic driving \cite{del2013shortcuts, li2017shortcuts, iram2021controlling, zhong2024time}. In the SM \cite{SM}, we demonstrate that these theoretical results also hold for a large class of active matter systems including the Active Ornstein-Uhlenbeck Process (AOUP) and the Active Brownian Particle (ABP) \cite{dabelow2019irreversibility, martin2021aoup}. 

{This Letter generalizes our previous result \cite{zhong2024beyond}, which dealt with the passive $\boldFnc = \boldzero$ case. The output work of a cyclic protocol in that passive setting would also satisfy Eq.~\eqref{eq:w_out_result}, but a gauge may be chosen (see \emph{Remark 2} below) where $A_i(\gamma) = 0$ everywhere \cite{statetostate}. In other words, in \cite{zhong2024beyond} we showed that conservative optimal control can be mapped exactly onto \emph{geometry}, whereas here, when a nonconservative $\boldFnc(\boldx)$ is introduced, the optimal control problem maps onto \emph{geometry}+\emph{electromagnetism}.}

This Letter is organized as follows: First, we derive Eq.~\eqref{eq:w_out_result}. We then discuss how to practically obtain optimal protocols when $U_\lambda(\boldx)$ is constrained. Following this, we translate theorems from classical EM into this optimal work extraction setting. We show that the housekeeping heat \cite{seifert2012stochastic, dechant2022geometric} plays the role of the electric potential, and generically may be made arbitrarily close to zero through strong confinement. Finally, we show that our theoretical results prove that the energetically efficient angle clamp protocol experimentally tested on $\mathrm{F_1}$-ATPase in \cite{mishima2025efficiently} is actually globally optimal, not only maximizing energy transduction but also minimizing heat dissipation simultaneously.

\emph{The Lorentz force Lagrangian}.---As a quick review of classical EM, a non-relativistic particle $\boldr(t) \in \mathbb{R}^3$ of mass $m$ and charge $q$ in a magnetic field describable by vector potential $\mathbf{A}(\boldr)$ obeys the Lorentz force law $m \ddot{\boldr} = q(\dot{\boldr} \times \mathbf{B})$, where $\mathbf{B} = \boldnabla \times \mathbf{A}$ is the magnetic field. More generally, the Lorentz force law may be derived as the Euler-Lagrange equation for the Lagrangian \cite{kibble2004classical}:
\begin{equation}
  L_\mathrm{EM}(\boldr, \dot{\boldr}) = \underbrace{  (m/2) \, \dot{r}^i \dot{r}^j \delta_{ij} }_{L_\mathrm{kinetic}}   + \underbrace{q \, \dot{r}^i  A_i(\boldr)}_{L_\mathrm{magnetic}} \label{eq:lorentz-lagrangian}.
\end{equation}
An elegant feature of the Lagrangian formalism is its generalizability to arbitrary geometries/dimensions by replacing $\delta_{ij}$ with the metric of the space embedding the particle $ \mathrm{g}_{ij}(\boldr)$, yielding the generalized Lorentz force law
\begin{equation}
     m\big( \mathrm{g}_{ij} \{ \ddot{r}^j + \Gamma_{a b}^j \,\dot{r}^a \dot{r}^b \} \big)     = q\big(  \dot{r}^k \{ \partial_i A_k - \partial_k A_i \} \big)  , \label{eq:general-Lorentz-force-law}
\end{equation}
where $\Gamma^{j}_{ab}$ are the Christoffel symbols for $\mathrm{g}_{ij}(\boldr)$ 
\cite{carroll2019spacetime}. 

\emph{Derivation.}---In this section we derive our main result [Eq.~\eqref{eq:w_out_result}]. Under the Langevin equation [Eq.~\eqref{eq:Langevin}], the thermodynamic state, which we identify with the ensemble probability density $\rho_t(\boldx)$, evolves according to
\begin{equation}
  \partial_t \rho_t = \mathcal{L}_{\lambda(t)}[\rho_t], \label{eq:rho-time-evolution}
\end{equation}
where
\begin{equation}
  \mathcal{L}_\lambda[\rho] := \boldnabla \cdot \{ \rho \, \mu (\boldnabla U_\lambda  - \boldFnc + \kBT \boldnabla \ln \rho) \} \label{eq:FP}
\end{equation}
is the Fokker-Planck operator \cite{linearFP}. For a given $\lambda$, there is a unique stationary-state distribution $\rhost_\lambda(\boldx)$ satisfying $\mathcal{L}_\lambda[\rhost_\lambda] = 0$ \cite{lyapunov}.
A defining characteristic of nonequilibrium stationary states is their non-zero probability current $\mathbf{J}^\mathrm{st}_\lambda(\boldx) = \rhost_{\lambda}(\boldx) \,\boldvst_\lambda(\boldx)$ satisfying $\boldnabla \cdot \mathbf{J}^\mathrm{st}_\lambda = 0$, where
\begin{equation}
    \boldvst_\lambda(\boldx) :=  -\mu \{ \boldnabla U_\lambda (\boldx) - \boldFnc(\boldx) + \kBT \boldnabla \ln \rhost_\lambda (\boldx)\} \label{eq:v-st}
\end{equation}
is the expected stationary-state velocity $\boldvst_\lambda(\boldx) = \langle\dot{\boldx} \, | \, \boldx\rangle^\mathrm{st}_{\lambda} $ \cite{hatano2001steady}. We assume a one-to-one mapping between stationary-state distributions and potential energy functions $ \rhost(\boldx) \leftrightarrow U(\boldx)$ \cite{Boltzmann}.

We now show that a particular decomposition of the protocol reveals the hidden Lagrangian structure to this optimal control problem. Given any $\tau$-periodic protocol $\lambda(t)|_{t \in [0, \tau]}$ yielding a $\tau$-periodic steady-state thermodynamic state $\rho_t|_{t \in [0, \tau]}$, we identify a {separate} $\tau$-periodic curve $\gamma(t)|_{t \in [0, \tau]} \in \mathcal{N} \supseteq \mathcal{M}$ in an appropriately expanded family of potential energy functions $\{ U_\gamma(\boldx) \}_{\gamma \in \mathcal{N}}$, such that the thermodynamic state is in instantaneous stationarity {with respect to $U_{\gamma(t)}$} at all $t$
\begin{equation}
    \rho_t(\boldx) = \rhost_{\gamma(t)}(\boldx). \label{eq:instantaneous-stationarity}
\end{equation}
This allows for the decomposition
\begin{equation}
    U_{\lambda(t)}(\boldx) = V_t(\boldx) + U_{\gamma(t)}(\boldx),  \label{eq:M-N-decomp}
\end{equation}
where the potential $V_t(\boldx)$ is defined by this equation, {such that the decomposition is valid for dynamics arbitrarily far from equilibrium.}

Eq.~\eqref{eq:M-N-decomp} is known as the Maes-Neto\u{c}n\'{y} decomposition \cite{maes2014nonequilibrium, dechant2022geometric}, which has the useful property that the time-evolution for $\rho_t = \rhost_{\gamma(t)}$ [Eq.~\eqref{eq:rho-time-evolution}] also satisfies
\begin{equation}
    \partial_t \rho_t = \boldnabla \cdot \{\rho_t (\mu\boldnabla {V}_t)\}, \label{eq:effective-CD}
\end{equation}
since the Fokker-Planck operator [Eq.~\eqref{eq:FP}] may be decomposed as $\mathcal{L}_{{\lambda(t)}}[\rho] = \boldnabla \cdot (\rho \, \mu \boldnabla V_t) + \mathcal{L}_{{\gamma(t)}} [\rho]$. Eq.~\eqref{eq:effective-CD} implies that $V_t(\boldx)$ is the counterdiabatic term that enforces Eq.~\eqref{eq:instantaneous-stationarity}  \cite{li2017shortcuts, zhong2024time}.

Using Eq.~\eqref{eq:M-N-decomp} and recalling that $\dot{\lambda}^i \{ \partial U_{\lambda(t)}/{\partial \lambda^i}\} =  {\partial_t U_{\lambda(t)}}$, the output work Eq.~\eqref{eq:work-out-definition} may be rewritten as
\begin{equation}
    -W_\mathrm{out} = \oint_0^\tau \bigg\{ \int \frac{\partial {V}_t}{\partial t} \,  \rho_t \dx\bigg\}  +  \bigg\{\int \frac{\partial U_{\gamma(t)} }{\partial t}  \rho_t \dx \bigg\}\dt. \label{eq:step1}
\end{equation}
By manipulating the first term via integrating-by-parts in $t$, applying Eq.~\eqref{eq:effective-CD}, and then integrating-by-parts in $\boldx$, we arrive at the expression
\begin{equation}
    -W_\mathrm{out} =  \oint_0^\tau \frac{1}{\mu}  \bigg\{\int \rhost_{\gamma(t)} |\mu\nabla {V}_t|^2 \dx \bigg\}  +  \dot{\gamma}^i \bigg\langle \frac{\partial U_{\gamma}}{\partial \gamma^i} \bigg\rangle^\st_{\gamma(t)}  \dt,  \label{eq:step2} 
\end{equation}
which may be written as 
\begin{equation}
   -W_\mathrm{out} =\oint_0^\tau \underbrace{\mu^{-1} \,\dot{\gamma}^i \dot{\gamma}^j g
_{i j} \big( \gamma(t) \big)}_{\dot{Q}_\mathrm{ex}} 
   + \underbrace{\dot{\gamma}^i A_{i} \big( \gamma(t) \big)}_{-\dot{W}_\mathrm{rev}}   \dt, \label{eq:work-out-expr1}
\end{equation}
where $g_{ij}(\gamma)$ is a positive-definite Riemannian metric defined on thermodynamic states $\gamma \in \mathcal{N}$ measuring $L^2$-optimal transport distances for the underlying steady state probability distributions \cite{zhong2024beyond, ito2024geometric, van2023thermodynamic}{---we give a brief review in the Appendix---}and $A_i(\gamma) = A_i^\mathrm{qs}(\gamma)$ is a thermodynamic one-form

\begin{equation}
   A^\mathrm{qs}_i(\gamma) =  \bigg\langle \frac{\partial U_{\gamma}}{\partial \gamma_i} \bigg\rangle^\mathrm{st}_\gamma \label{eq:Aqs} 
\end{equation}
that is the \emph{quasistatic} work cost to vary $\gamma_i \rightarrow \gamma_i + \delta \gamma_i$ {\cite{footnote-total-derivative}}. We label the two terms as the excess heat $\dot{Q}_\mathrm{ex}$ (which is always nonnegative) and the negative reversible work $\dot{W}_\mathrm{rev}$, with sign convention so that $W_\mathrm{out} = W_\mathrm{rev} - Q_\mathrm{ex}$. 

We have arrived at our main result: the integrand of Eq~\eqref{eq:work-out-expr1} is a generalized Lorentz force Lagrangian defined on the space of thermodynamic states ({\it i.e.},  parametric stationary-state probability distributions $\mathcal{P}^\mathrm{st}_{\mathcal{N}}(\mathbb{R}^d) := \{\rhost_\gamma\}_{\gamma \in \mathcal{N}}$). In order for a curve in thermodynamic state space $\gamma^*(t) \in \mathcal{N}$ to be a minimum of Eq~\eqref{eq:work-out-expr1}, it must obey the Lorentz force law {[Eq.~\eqref{eq:general-Lorentz-force-law}]} under the metric tensor $g_{ij}(\gamma)$, magnetic potential $A_i(\gamma)$, and charge-to-mass ratio $\mu/2$. We wish to emphasize that this result is valid arbitrarily far from equilibrium; revealing this \emph{exact} structure constitutes our first major contribution.

\emph{Computing optimal protocols.}---The space of possible thermodynamic state trajectories is subject to the constraint of limited control expressivity \cite{controlexpressivity}. Nonetheless, Eq.~\eqref{eq:work-out-expr1} tells us that limited-control optimal protocols should still realize thermodynamic state trajectories that obey the thermodynamic Lorentz force law when subjected to a limited control constraint embedded in an, {\it e.g.}, Pontryagin manner \cite{zhong2022limited}. 

Assuming that stationary-state expectations can be measured (so that we can estimate $g_{ij}(\gamma)$ and $A_i(\gamma)$ for $\gamma \in \mathcal{M}$), Eq.~\eqref{eq:work-out-expr1} allows us to compute near-optimal (if not optimal) maximal work protocols, akin to Alg.~1 in \cite{zhong2024beyond}: First, variationally optimize the curve $\gamma(t) \in \mathcal{M}$ that minimizes Eq.~\eqref{eq:work-out-expr1}. Once an optimal path $\gamma^*(t)$ has been identified, an optimal protocol $\lambda^*(t)$ that endeavors to implement the desired thermodynamic state trajectory $\rho_t = \rhost_{\gamma^*(t)}$ can be obtained through the inclusion of a geometrically calculable counterdiabatic driving term; see the SM for further details \cite{SM}.

Given our main result, well-known properties of classical EM give rise to consequences for globally optimal nonequilibrium engines:

\emph{Remark 1: Conserved quantity}.---For the Lorentz Lagrangian in the integrand of Eq.~\eqref{eq:work-out-expr1}, the quantity $E := \dot{\gamma}^i (\partial L / \partial \dot{\gamma}^i) - L = \mu^{-1} \dot{\gamma}^i \dot{\gamma}^j g_{ij}(\gamma) = \dot{Q}_\mathrm{ex}$ is conserved in optimal trajectories $\gamma^*(t)$. In order for optimal engines to maximize output work, excess heat must be dissipated at a constant rate \cite{KE}, consistent with previous results for slow processes  \cite{diosi1996thermodynamic, sivak2012thermodynamic,Brander_PRL_engines,frim2022geometric}---however, our result is exact for protocols of any duration.

\emph{Remark 2: Gauge Invariance}.---Similar to classical EM, there is a gauge invariance $A_i(\gamma) \rightarrow A_i(\gamma) + \partial \Psi (\gamma) / \partial \gamma^i$ for thermodynamic state space scalar fields $\Psi(\gamma)$
. In particular, the gauge
\begin{equation}
    A^\mathrm{nc}_i(\gamma) =  A^\mathrm{qs}_i(\gamma) - {\partial \mathcal{F}^\mathrm{st}(\gamma)}/{\partial \gamma^i},
\end{equation}
where
\begin{equation}
    \mathcal{F}^\mathrm{st}(\gamma) = \int \rhost_\gamma(\boldx) \{ U_\gamma(\boldx) + \kBT \,\ln \rhost_\gamma(\boldx)\}\dx 
\end{equation}
is the nonequilibrium stationary state free energy functional \cite{rosa2024variational}, may be shown to satisfy 
\begin{equation}
  \dot{\gamma}^i A_i^\mathrm{nc}(\gamma) = \int \rhost_{\gamma(t)} (\mu \boldnabla V_t \cdot \boldFnc ) \dx, \label{eq:Anc}
\end{equation}
which is invariant to shifts by spatially uniform offsets of the potential $U_\gamma(\boldx) \rightarrow U_\gamma(\boldx) + \Psi(\gamma)$. Whereas $A^\mathrm{qs}_i$ [Eq.~\eqref{eq:Aqs}] is evaluable as a stationary state expectation, $A_i^\mathrm{nc}$ is useful for later derivations. {It is also readily seen that $A^\mathrm{nc}_i(\gamma) = 0$ everywhere when $\boldFnc = \boldzero$.}

\emph{Remark 3: Housekeeping heat is analogous to the electric potential}.---It is also straightforward to show using Eq.~\eqref{eq:M-N-decomp} that the work done by the nonconservative-forcing $W_\mathrm{nc} = \oint_0^\tau \big\langle \boldFnc \circ \dot{\boldx} \big\rangle \dt$ (here $\circ$ denotes a Stratonovich product) may be written as 
\begin{align}
  W_\mathrm{nc} &= \oint_0^\tau   { \underbrace{ \sigma_\mathrm{hk}\big( \gamma(t) \big)}_{\dot{Q}_\mathrm{hk}}  } + \underbrace{- { \dot{\gamma}^i A_i \big( \gamma(t) \big)}}_{\dot{W}_\mathrm{rev}} \dt, \label{eq:input-work}
\end{align}
where $\sigma_\mathrm{hk} = \big\langle \boldFnc \circ \dot{\boldx} \big\rangle^\st_\gamma$ is the 
housekeeping-heat rate \cite{seifert2012stochastic, dechant2022geometric}
\begin{equation} 
  \sigma_\mathrm{hk}(\gamma) = \mu^{-1} \int \rhost_\gamma(\boldx) \,|\boldvst_\gamma(\boldx)|^2 \dx \label{eq:housekeeping-EP}
\end{equation} 
{defined with $\boldvst_\gamma(\boldx)$ [Eq.~\eqref{eq:v-st}].} 

Per the first law of thermodynamics $W_\mathrm{nc} = W_\mathrm{out} + Q_\mathrm{out}$, the output heat has the exact expression
\begin{equation}
    Q_\mathrm{out} = \oint_0^\tau \underbrace{\mu^{-1} \, \dot{\gamma}^i \dot{\gamma}^j g
_{i j} \big( \gamma(t) \big)}_{\dot{Q}_\mathrm{ex}} + \underbrace{\sigma_\mathrm{hk} \big( \gamma(t)\big)}_{\dot{Q}_\mathrm{hk}} \dt. 
\label{eq:q-out}
\end{equation}
This has the form of a kinetic ($\dot{\gamma}^2$-dependent) term and a potential (state-dependent) term. In terms of EM, $\sigma_\mathrm{hk}(\gamma)$ corresponds to $L_\mathrm{electric} = -q\Phi(\boldr)$ where $\Phi(\boldr)$ is the electric potential. We note that this observation (viewing $\sigma_\mathrm{hk}$ as the negative potential energy) has been made before \cite{bo2013optimal, davis2024active}. 

Altogether, the flow of energy between $W_\mathrm{nc}$, $W_\mathrm{out}$, and $Q_\mathrm{out}$, and the exact relationship to EM, are depicted in the following diagram:
\begin{equation} 
\centering\begin{tikzcd}
{W}_\mathrm{nc} 
\arrow["\oint_0^\tau \dot{Q}_\mathrm{hk} \dt", dr, swap] 
\arrow["\oint_0^\tau \dot{W}_\mathrm{rev} \dt",rr]
  & 
    &{W}_\mathrm{out}  
     \arrow[dl, "\oint_0^\tau \dot{Q}_\mathrm{ex} \dt"] \\ [1ex]
  & {Q}_\mathrm{out} 
\end{tikzcd} \nonumber
\end{equation}
\begin{equation}
\begin{tabular}{ c | c } 
    Noneq. engine for $\rho_t = \rhost_{\gamma(t)}$ \ \  & EM for $\boldr(t)$ \\ 
    \hline
  $\dot{Q}_\mathrm{ex} = \mu^{-1} \, \dot{\gamma}^i \dot{\gamma}^j g_{ij}(\gamma)$ & $L_\mathrm{kinetic} = m |\dot{\boldr}|^2 / 2 $ \\
  $\dot{W}_\mathrm{rev} = -\dot{\gamma}^i A_i (\gamma) $& \ \ $L_\mathrm{magnetic} = q \dot{\boldr} \cdot \mathbf{A}(\boldr) \ $ \\ 
  $\dot{Q}_\mathrm{hk} = \sigma_\mathrm{hk}(\gamma)$ & $L_\mathrm{electric} = - q \Phi(\boldr). $ 
\end{tabular}  \nonumber
\end{equation}
Depending on what nonequilibrium thermodynamic quantity one wishes to optimize ({\it e.g.}, the work extraction $W_\mathrm{out}$, total heat out $Q_\mathrm{out}$, any linear combination, etc.), the objective function is a Lorentz force Lagrangian action for $\gamma(t)$, constructed from these three components alone;  under this exact analogy between controlled nonconservative systems and EM, a thermodynamically optimal protocol is one that drives the thermodynamic state to satisfy a Lorentz force law.

\emph{Zero housekeeping heat for tight confinement.}---For potentials of the form
\begin{equation}
  U_\gamma(\boldx) = U_\mathrm{fixed}(\boldx) + \kappa_\gamma |\boldx - \boldr_\gamma |^2 / 2, \label{eq:potential-form}
\end{equation}
where $\gamma = (\kappa_\gamma, \boldr_\gamma)$, we show in the {SM} \cite{SM} that 
\begin{equation}
    \lim_{\kappa_\gamma \rightarrow \infty} \sigma_\mathrm{hk}(\gamma) = 0.  \label{eq:zero-hk-heat}
\end{equation}
with convergence rate $\sigma_\mathrm{hk}(\gamma) \sim \kappa_\gamma^{-1}$. {Thus, protocols that confine the system tightly in configuration space produce zero housekeeping entropy production; all dissipation is from excess heat alone.}

{\emph{Transitions between stationary states}.---We remark that if instead the protocol were a finite-time transition between stationary states (as considered in, {\it e.g.},} \cite{davis2024active, casert2024learning, zulkowski2013optimal, monter2025optimal}) {with specified boundary conditions $\lambda(0) = \lambda_i$, $\lambda(\tau) = \lambda_f$, and $\rho_0 = \rhost_{\lambda_i}$, the only modification to our derivation is that the integration-by-parts in $t$ performed between Eq.~\eqref{eq:step1} and Eq.~\eqref{eq:step2} accrues a boundary term, yielding}
\begin{gather}
    -W_\mathrm{out}^{\lambda_{i \rightarrow}\lambda_f} = \int_0^\tau \bigg[\frac{\dot{\gamma}^i \dot{\gamma}^j  }{\mu}  g_{i j}+ \dot{\gamma}^i A_{i}^\mathrm{qs} \bigg] \mathrm{d}t + \big\langle V_t \big\rangle \bigg|^\tau_{t = 0} \nonumber \\ 
= \int_0^\tau \bigg[\frac{\dot{\gamma}^i \dot{\gamma}^j  }{\mu}  g_{i j}+ \dot{\gamma}^i A_{i}^\mathrm{nc} \bigg] \mathrm{d}t + \langle U_{\lambda_f} \rangle^\st_{\gamma(\tau)}  - \langle U_{\lambda_i} \rangle^\st_{\gamma(0)}.
\label{eq:W_transit}
\end{gather}
{Eq.~\eqref{eq:W_transit} indicates that for state-to-state transitions, one should perfom optimization over curves $\gamma(t)|_{t\in[0,\tau]}$, with $\gamma(0) = \lambda_i$ and without a terminal constraint for $\gamma(\tau)$  \cite{zhong2024beyond}. Discontinuous jumps for $\lambda^*(t)$ at $t = 0$ and $t = \tau$ have been observed \cite{casert2024learning} similar to the $\boldFnc = \boldzero$ case \cite{schmiedl2007optimal}, but can be explained by the decomposition Eq.~\eqref{eq:M-N-decomp} \cite{zhong2024beyond}.}

\emph{Example: ATP-synthase}.---\cite{mishima2025efficiently} recently studied experimental protocols rotating the $F_1$ protein that makes up ATP-synthase, showing the thermodynamic superiority of a constant-velocity angle clamp protocol. We note that they applied positive work into the nonconservative forcing
(so intead, $W_\mathrm{out} \rightarrow W_\mathrm{in}$ and flipping the arrow for $\dot{W}_\mathrm{rev}$ in our diagram). See \cite{gupta2022optimal, wareham2024multi} for linear response treatments.

We will prove that their constant-velocity angle clamp protocol implements the \emph{globally} optimal active engine, and simultaneously maximizes work transduction and minimizes heat dissipation. Here we consider a periodic one-dimensional configuration space $\theta \in [0, 2\pi)$ that undergoes the nonconservative Langevin equation 
\begin{equation}
     \dot{\theta} = - \mu \,\partial_\theta U_{\lambda(t)} + \mu\Fnc  + \sqrt{2 \mu \kBT }\, \xi(t),
\end{equation}
where now $\Fnc $ is a scalar nonconservative torque, and $\xi(t)$ is a scalar delta-correlated Gaussian white noise. $\Fnc$ directly corresponds to a chemical potential that drives a nonconservative torque, and the protocol can take on the generic form for $\lambda = (\kappa_\lambda, \varphi_\lambda)$:
\begin{equation}
  U_{\lambda}(\theta) = u_\mathrm{fixed}(\theta) +  \kappa_\lambda [ 1 - \cos(\theta - \varphi_\lambda)]. \label{eq:atp-protocol-form}
\end{equation}

We first derive the globally optimal thermodynamic trajectory $\rho_t^*(\theta)$: we consider a thermodynamic state manifold $\gamma \in \mathcal{N}$, where the angular coordinate $\alpha$ may be identified so that for $\gamma = (\alpha, \gammaperp)$,
\begin{equation}
  \rhost_{(\alpha, \Sigma)}(\theta) = \rhost_{ \Sigma}(\theta - \alpha). 
\end{equation}
Here we have denoted $\rhost_{ \gammaperp} := \rhost_{(0, \gammaperp)} $. The variables $\gammaperp$ parameterize the shape of the distribution $\rhost_\gamma(\theta)$, while $\alpha$ parametrizes its angular translation. Coordinates for $\Sigma$ may be chosen via the Gram-Schmidt procedure to be orthogonal to $\alpha$; {\it i.e.}, $g_{\alpha \gammaperp} = 0$ for all $\gamma \in \mathcal{N}$. 

We analytically derive in the SM \cite{SM} that the thermodynamic metric $g_{ij}(\gamma)$, one-form $A_i(\gamma) = A_i^\mathrm{nc}(\gamma)$ [Eq.~\eqref{eq:Anc}], and housekeeping heat $\sigma_\mathrm{hk}(\gamma)$ satisfy
\begin{equation}
    g_{\alpha \alpha} = C ,   \ \ A_\alpha = -C \Fnc, \ \ \sigma_\mathrm{hk} =  (1 - C)\Fncsquared; \label{eq:particle-on-ring-thermo-geo}
\end{equation}
where 
\begin{equation}
  C(\gammaperp) = 1 - \frac{4\pi^2}{ \int_0^{2\pi}  \{ \rhost_{\gammaperp}(\theta) \} ^{-1}   \dvarphi}  \in [0, 1] \label{eq:concentration-constant}
\end{equation}
is an $\alpha$-independent ``concentration constant'' that varies between $C = 0$ for a flat distribution $\rhost_{\gammaperp}(\theta) = 1/2\pi$, to $C = 1$ for, {\it e.g.}, any Dirac delta distribution $\rhost_{\gammaperp}(\theta) = \delta(\theta - \theta_0)/2\pi$ \cite{fullsupport}. It may now be seen that, consistent with Eq.~\eqref{eq:zero-hk-heat}, the housekeeping heat may be made arbitrarily close to zero by having $C(\Sigma)$ approach $1$; given a finite $\Fnc$, this may be achieved by an infinite stiff and narrow trap so that $\rhost_\Sigma(\theta)$ is a Dirac delta distribution. 

Choosing orthogonal coordinates $g_{\alpha \Sigma} = 0$ implies $A_\Sigma^\mathrm{nc} = 0$ at all $\gamma \in \mathcal{N}$ (see SM \cite{SM}). In these coordinates, because $A_\Sigma = 0$ and $g_{\Sigma \Sigma} > 0$ (due to positivity of the metric), $\Sigma(t)$ must be fixed under optimality; otherwise it would contribute a positive $\dot{Q}_\mathrm{ex}$ cost without any $\dot{W}_\mathrm{rev}$ reward. Thus, under optimality, only $\alpha(t)$ is time-dependent, so the output work is 
\begin{equation}
  -W_\mathrm{out}[\alpha(t)] = C(\Sigma_0) \oint_0^\tau\big\{  \dot{\alpha}^2 /\mu - \mu F_\mathrm{nc} \dot{\alpha}\big\} \dt.
\end{equation}
The Euler-Lagrange equation is $\ddot{\alpha}^* / \mu = 0$ for optimal trajectories, yielding $\dot{\alpha}^* = \mathrm{const}$. 
Thus, optimal-work state trajectories $\gamma^*(t) \in \mathcal{N}$ are of the form
\begin{equation}
    \alpha^*(t) = \alpha_0 +  \omega t \quad \mathrm{and} \quad \Sigma^*(t) = \Sigma_0,
\end{equation}
which, with Eq.~\eqref{eq:particle-on-ring-thermo-geo}, yields the diagram
\[
\begin{tikzcd}
\dot{W}_\mathrm{nc}       \arrow["\dot{Q}_\mathrm{hk} = (1 - C_0)  \mu \Fncsquared  ", dr, swap] 
\arrow["\dot{W}_\mathrm{rev} = C_0 \omega \Fnc",rr]
  &     
    & \dot{W}_\mathrm{out} \arrow[dl, "\dot{Q}_\mathrm{ex} = C_0 \omega^2  /\mu "] \\
  & \dot{Q}_\mathrm{out} 
\end{tikzcd}
\]
where $C_0 = C(\Sigma_0)$. Because all these expressions are linear in $C_0$, optima are achieved either at the limits $C_0 = 0$ (with $\rhost_\gamma(\theta) = 1/2\pi$) yielding constant housekeeping heat $\dot{Q}_\mathrm{hk} = \mu \Fncsquared$ but no output work; or $C_0 = 1$ which yields $\dot{Q}_\mathrm{hk} = 0$, and $\omega$-dependent $\dot{W}_\mathrm{nc}$ and $\dot{W}_\mathrm{out}$. The output power $\dot{W}_\mathrm{out} = C_0\omega(\Fnc - \omega /\mu)$ is maximized at $C_0^* = 1$ and $\omega^* = \mu \Fnc / 2$, resulting in
\begin{equation}
    \dot{W}_\mathrm{out}^* = \mu \Fncsquared / 4 \quad \mathrm{from} \quad \dot{W}_\mathrm{nc}^* = \mu \Fncsquared / 2, \label{eq:optimal-output-power}
\end{equation}
which reproduces the well-known result of $\dot{W}_\mathrm{out} / \dot{W}_\mathrm{nc}= 1/2$ efficiency at maximum power \cite{van2012efficiency, reducedfq}. 

Finally, globally optimal solutions [$C(\Sigma_0) = 1$ and $\alpha(t) = \alpha_0 + \omega t$] can be realized by the protocol $\lambda^*(t) = (\kappa_\lambda \rightarrow \infty, \varphi_{\lambda(t)} = \alpha_0 + \omega t)$ for Eq.~\eqref{eq:atp-protocol-form} \cite{blaber2022optimal}. Remarkably, this is exactly the constant-angular clamp protocol implemented in \cite{mishima2025efficiently}. Furthermore, our theoretical results provide predictions for new experiments under the setup in \cite{mishima2025efficiently}: for any protocol, work measurements should exactly match our expressions Eqs.~\eqref{eq:work-out-expr1},~\eqref{eq:particle-on-ring-thermo-geo}, and~\eqref{eq:concentration-constant}. 

\emph{Concluding Remarks.}---It has been observed that for variable-temperature systems \cite{frim2022geometric, miangolarra2022geometry}, the work extraction can be written as a squared thermodynamic length cost plus a thermodynamic area reward. We remark that their objective functions may be recast as a Lorentz force Lagrangian action [Eq.~\eqref{eq:work-out-expr1}], as Stokes' Theorem allows for bounded area integrals to be expressed as a line-integral  \cite{purcell2013electricity}. Line-integrals have the practical advantage of being much easier to evaluate. 

As mentioned in the Introduction, our results are immediately applicable to the AOUP and ABP (see SM \cite{SM}). It should be straightforward to extend our results to underdamped \cite{sabbagh2024wasserstein} and discrete-state \cite{van2023thermodynamic} settings, where passive dissipation has been shown to be optimal-transport-geometric. It would be exciting to apply our theory to high-dimensional active systems and field theories \cite{davis2024active, soriani2025control}, and active quantum systems \cite{lacerda2025information} where quantum coherence may be exploitable as a resource \cite{shi2020quantum}. It remains to be seen how this theory of externalized optimal control is applicable to actual autonomously-functioning biological motors~\cite{pietzonka2019autonomous, lathouwers2020nonequilibrium}, but we expect that our results---applicable arbitrarily far from equilibrium---will provide insight into the optimal design and control of nonequilibium nanoscopic systems. 

\emph{Acknolwedgements.}---We thank Étienne Fodor, Ben Kuznets-Speck, Hana Mir, and David Sivak for insightful conversations. 
A.Z. was supported by the Department of Defense (DoD) through the National Defense Science \& Engineering Graduate (NDSEG) Fellowship Program. A. G. F. was supported by the NSF (DMR-2005749) and through the NSF GRFP (DGE-1752814). This research was supported by CoCoSys, one of the seven centers in JUMP 2.0, a Semiconductor Research Corporation (SRC) program sponsed by DARPA; in part by grants from the NSF (DMS-2235451) and Simons Foundation (MPS-NITMB-00005320) to the NSF-Simons National Institute for Theory and Mathematics in Biology (NITMB); as well as in part by the U.S. Army Research Laboratory and the U.S. Army Research Office under Contract No. W911NF-20-1-0151.

\bibliography{main}

\appendix
\onecolumngrid
\section*{End Matter}
\twocolumngrid

{\emph{Appendix: The metric tensor }$g_{ij}$.}{---We showed in \cite{zhong2024beyond} that the optimal transport metric tensor \cite{otto2001geometry}, defined as}
\begin{equation}
    g_{ij}(\gamma) = \int \rhost_\gamma(\boldx) \boldnabla \phi_i(\boldx) \cdot \boldnabla \phi_j(\boldx) \dx, 
\end{equation}
{where $\phi_i(\boldx)$ is the unique scalar field satisfying the continuity equation}
\begin{equation}
    \frac{\partial \rhost_\gamma(\boldx)}{\partial \gamma^i} = \boldnabla \cdot [\rhost_\gamma (\boldx) \boldnabla \phi_i(\boldx)],
\end{equation}
{is \emph{equivalent} to the Sivak and Crooks metric tensor \cite{sivak2012thermodynamic} that is defined in terms of a time-correlation function}
\begin{equation}
   g_{ij}(\gamma) = (\mu \kBT)^{-1}  \int_0^\infty \big\langle \delta f^\mathrm{eq}_i\big( \boldx(t') \big) \, \delta f^\mathrm{eq}_j\big( \boldx(0) \big)  \big\rangle ^\mathrm{eq}_\gamma \dt' \label{eq:sivak-crooks}
\end{equation}
{under the effective \emph{equilibrium} dynamics }
\begin{equation}
    \dot{\boldx} = - \mu \boldnabla U^\mathrm{eq}_\gamma(\boldx)  + \sqrt{2 \mu\kBT } \, \boldxi(t) \label{eq:Langevin-equilibrium}
\end{equation}
{initialized in steady state $\boldx(0) \sim \rhost(\cdot)$. Here $U^\mathrm{eq}_\gamma(\boldx) = -\ln \rhost_\gamma(\boldx)$ is the effective equilibrium potential energy whose Boltzmann distribution matches the nonequilibrium stationary state distribution, and $\delta f^\mathrm{eq}_i(\boldx) = -\partial_{\gamma^i} U^\mathrm{eq}_\gamma ( \boldx ) + \langle \partial_{\gamma^i} U^\mathrm{eq}_\gamma \rangle^\mathrm{st}_\gamma$ is the so-called \emph{excess conjugate force} that (crucial for our prior results) satisfies $\delta f^\mathrm{eq}_i(\boldx) = \kBT \partial_{\gamma^i} \ln \rhost_\gamma(\boldx)$. Typically, $U^\mathrm{eq}_\gamma (\boldx) \neq U_\gamma(\boldx)$ when $\boldFnc(\boldx) \neq \boldzero$. 

By comparing Eqs.~(11) and (13) to Eqs.~(35) and (36), one can relate $\mu V_t(\boldx) = \dot{\gamma}^i \phi_i(\boldx)$, which allows the first term in Eq.~(13) to be written in terms of the optimal transport thermodynamic metric tensor Eq.~(14). 

A difficulty in our nonconservative setting [Eq.~\eqref{eq:Langevin}] is that in order to calculate $g_{ij}(\gamma)$, one must know  $U^\mathrm{eq}_\gamma (\boldx)  = -\kBT  \ln \rhost_\gamma(\boldx)$ in order to use Eqs.~\eqref{eq:sivak-crooks} and~\eqref{eq:Langevin-equilibrium} to calculate the metric tensor, which may be learned using modern machine learning algorithms (see, e.g., \cite{boffi2024deep}). However $g_{ij}(\gamma)$ may also be approximated by symmetrizing the linear-response tensor }
\begin{equation}
    g_{ij} \approx (g^\mathrm{lr}_{ij} + g^\mathrm{lr}_{ji})/2
\end{equation}
{where $ g^{\mathrm{lr}}_{ij} = (\mu \kBT)^{-1}  \int_0^\infty \big\langle \delta f_i\big( \boldx(t') \big) \, \delta f_j\big( \boldx(0) \big)  \big\rangle ^\mathrm{st}_\gamma \dt'$ is defined under the nonconservative Langevin dynamics [Eq.~\eqref{eq:Langevin}] with $\delta f_i ( \boldx )  =  -\partial_{\gamma^i} U_\gamma ( \boldx ) + \langle \partial_{\gamma^i} U_\gamma \rangle^\mathrm{st}_\gamma$ readily evaluable. It may be shown via perturbation theory that this approximation is accurate up to second order $o(|\boldFnc|^2)$.}

\end{document}


\newcommand{\boldzeta}{{\boldsymbol{\zeta}}}
\newcommand{\boldlambda}{{\boldsymbol{\lambda}}}
\newcommand{\boldx}{{{x}}}
\newcommand{\boldq}{{\boldsymbol{q}}}

\newcommand{\trajx}{X}
\newcommand{\boldc}{c}
\newcommand{\dt}{{\,\mathrm{d}t}}
\newcommand{\ds}{{\,\mathrm{d}s}}
\newcommand{\dx}{{\,\mathrm{d}x}}
\newcommand{\dee}{{\mathrm{d}}}
\newcommand{\gammadot}{{\dot{\gamma}}}

\renewcommand{\st}{{\mathrm{st}}}

\newcommand{\boldrho}{{\boldsymbol{\rho}}}
\newcommand{\boldU}{{\boldsymbol{U}}}
\newcommand{\boldpi}{{\boldsymbol{\pi}}}
\newcommand{\boldpsi}{{\boldsymbol{\psi}}}
\newcommand{\curlyL}{{\cal{L}}}
\newcommand{\curlyW}{{\cal{W}}}

\newcommand{\tee}{{\mathrm{t}}}
\newcommand{\tf}{{\mathrm{t_f}}}
\newcommand{\A}{{\mathrm{A}}}
\newcommand{\B}{{\mathrm{B}}}
\newcommand{\C}{{\mathrm{C}}}
\newcommand{\D}{{\mathrm{D}}}
\newcommand{\U}{{\mathrm{U}}}
\newcommand{\W}{{\mathrm{W}}}
\newcommand{\bolddx}{{\Delta \boldsymbol{x}}}

\newcommand{\kBT}{{k_\mathrm{B} T}}
\newcommand{\vecF}{\boldsymbol{f}_\mathrm{nc}}
\newcommand{\veczeta}{\boldzeta}

\renewcommand{\st}{{\mathrm{st}}}
\newcommand{\rhost}{\rho^\mathrm{st}} 
\newcommand{\boldvst}{\boldsymbol{v}^\mathrm{st}} 
\newcommand{\vst}{{v}^\mathrm{st}} 
\newcommand{\boldFzero}{\boldsymbol{F}_0}

\newcommand{\boldA}{\boldsymbol{A}}
\newcommand{\boldr}{\mathbf{r}}
\newcommand{\Lgamma}{\mathcal{L}_\gamma}

\newcommand{\boldxi}{{\boldsymbol{\xi}}}
\newcommand{\boldv}{{\boldsymbol{v}}} 
\renewcommand{\boldx}{{\boldsymbol{x}}}
\newcommand{\dvarphi}{{\,\mathrm{d}\theta}}
\newcommand{\boldnabla}{{\boldsymbol{\nabla}}}

\newcommand{\boldFnc}{{\boldsymbol{F}_\mathrm{nc}}}
\newcommand{\Fnc}{{{F}_\mathrm{nc}}}
\newcommand{\Fncsquared}{{{F}_\mathrm{nc}^2}}
\newcommand{\Fncinverse}{{{F}_\mathrm{nc}^{-1}}}
\newcommand{\boldzero}{{\boldsymbol{0}}}

\preprint{APS/123-QED}

\title{Supplementary Information for ``Optimal active engines obey the thermodynamic Lorentz force law''
}
\author{Adrianne Zhong$^{1,2,3}$}
\email{adrianne.zhong@northwestern.edu}
\author{Adam G. Frim$^{2,4}$}
\author{Michael R. DeWeese$^{2,3,5}$}
\affiliation{%
 ${^1}$NSF-Simons National Institute for Theory and Mathematics in Biology, Chicago, IL, 60611\\
 ${^2}$Department of Physics, University of California, Berkeley, Berkeley, CA, 94720 \\
${^3}$Redwood Center For Theoretical Neuroscience, University of California, Berkeley, Berkeley, CA, 94720 \\
  ${^4}$Department of Physics and Astronomy and the Center for Soft and Living Matter, University of Pennsylvania, Philadelphia, PA, 19104 \\
${^5}$Department of Neuroscience, University of California, Berkeley, Berkeley, CA, 94720
}%

\date{\today}
\maketitle
\section{Mapping active Gaussian colored noise to our nonconservative setting.}

Here we show that active overdamped particles driven by Gaussian colored noise (as with the Active Ornstein-Uhlenbeck Process and the Active Brownian Particle \cite{dabelow2019irreversibility, martin2021aoup}) fit within our formalism. We shall concern ourselves with a single $d = 1$ active particle, as the generalization to an arbitrary number of particles and dimensions immediately follows. The Langevin equation for this single active particle with position $x \in \mathbb{R}$ is: 
\begin{equation}
    \dot{x} = - \mu \partial_x \Phi_\lambda (x) + v(t) + \sqrt{2 \mu \kBT } \,\xi_x(t), \label{eq:active-langevin-x}
\end{equation}
where the self-propulsion term $v(t)$ has the statistics $\langle v(t)\rangle = 0$ and $\langle v(t) v(t')\rangle = 2 (D_\mathrm{active} / \tau_\mathrm{active}) \, e^{-|t - t'|/\tau_\mathrm{active}}$, and $\xi_x(t)$ is standard Gaussian white noise with $\langle \xi_x(t)\rangle =0$ and $\langle \xi_x(t) \xi_x(t')\rangle = \delta(t - t')$.
The physically measurable protocol output work is 
\begin{equation}
    W_\mathrm{out}[\lambda(t)] = -\oint_0^\tau \dot{\lambda}^i \bigg\langle \frac{\partial \Phi_\lambda (x)}{\partial \lambda^i} \bigg\rangle \dt . 
\end{equation}
The colored noise $v(t)$ can itself be generated from an Ornstein-Uhlenbeck process
\begin{equation}
    \tau_\mathrm{active}\, \dot{v} =  - v + \sqrt{2D_\mathrm{active}} \, \xi_\mathrm{active}(t). \label{eq:active-langevin-v}
\end{equation}
where $\xi_\mathrm{active}(t)$ is another standard Gaussian white noise with unit magnitude  $\langle \xi_\mathrm{active}(t)\rangle =0$ and $\langle \xi_\mathrm{active}(t) \,\xi_\mathrm{active}(t')\rangle = \delta(t - t')$ \cite{fodor2016far}.

By performing the coordinate transformation $y = ( \mu \kBT / D_\mathrm{active}) \tau_\mathrm{active} v $, we can rewrite Eqs.~\eqref{eq:active-langevin-x} and \eqref{eq:active-langevin-v} into a single composite Langevin equation with uniform Gaussian white noise for $\boldx = (x, y) \in \mathbb{R}^2$:
\begin{equation}
    \dot{\boldx} = -\mu \boldnabla U_\lambda(\boldx) + \mu \boldFnc (\boldx) + \sqrt{2 \mu \kBT} \, \boldxi (t) 
\end{equation}
where
\begin{align}
    U_\lambda(x, y) = \Phi_\lambda (x) + k y^2 / 2,  \quad\quad \boldFnc(x, y) &= \frac{D_\mathrm{active}}{\mu \kBT } \begin{bmatrix}
           k y  \\
           0
         \end{bmatrix}, \quad \quad \mathrm{and} \quad \quad \boldxi(t) = \begin{bmatrix}
           \xi_x(t) \\
           \xi_\mathrm{active}(t)
         \end{bmatrix}, 
\end{align}
with $ k = (\mu \tau_\mathrm{active})^{-1}$. Clearly, $\boldFnc(x, y)$ satisfies $\boldnabla \cdot \boldFnc = (D_\mathrm{active} /\mu\kBT) \{ \partial_x (ky) + \partial_y 0 \} = 0. $

Thus, we have mapped the $d = 1$ active overdamped Langevin equation via a time-correlated self-propulsion, into a $d = 2$ nonconservatively-driven overdamped Langevin equation with isotropic standard Gaussian white noise. It is straightforward to extend this derivation to multidimensional and many-body active systems. Likewise, our derivation may be extended to the active Brownian particle [where instead, $v(t) = v_0 \cos \theta(t)$ with $\ \dot{\theta}(t) = \sqrt{2D / v_0 \tau_\mathrm{active}} \,\eta_\mathrm{active}(t)$].

The physically measurable output work is $W_\mathrm{out} = -\oint_0^\tau \dot{\lambda}^i \langle \partial_{\lambda^i} \Phi_\lambda (x)\rangle \dt = -\oint_0^\tau \dot{\lambda}^i \langle \partial_{\lambda^i} U_\lambda (x, y)\rangle \dt$, so our relation
\begin{equation}
    -W_\mathrm{out} = \oint_0^\tau \bigg[ \mu^{-1} \,\dot{\gamma}^i \dot{\gamma}^j g_{ij}(\gamma(t))\,  + \dot{\gamma}^i A_i(\gamma(t))  \bigg] \mathrm{d}t
\end{equation}
for the joint thermodynamic state $\rho_t(x, y) = \rhost_{\gamma(t)}(x, y)$ is valid, thus demonstrating the applicability of our formalism to active, colored-noise-driven Langevin systems. 

\section{Engineering Optimal Protocols}

Here we discuss how to practically compute (near-)optimal protocols given limited access to measuring stationary-state expectations for $\lambda \in \mathcal{M}$. We are assuming the linear (i.e., control-affine) protocol parameterization 
\begin{equation}
    U_\lambda(\boldx) = U_\mathrm{fixed}(\boldx) + \lambda^i U_i(\boldx). 
\end{equation}
This allows us to write the decomposition $U_{\lambda(t)}(\boldx) = U_{\gamma(t)}(\boldx) + V_t(\boldx)$ purely in terms of the protocol parameters themselves: $\lambda(t) = \gamma(t) + \eta(t)$ for 
\begin{equation}
    U_\gamma(\boldx) = U_\mathrm{fixed}(\boldx) + \gamma^i U_i(\boldx) \quad\quad\quad\mathrm{and}\ \ \quad\quad V_t(\boldx) = \eta^i U_i(\boldx).
\end{equation}
To outline our procedure:

\textbf{A.} Solve the optimal $\tau$-periodic $\gamma^*(t) \in \mathcal{M}$ that minimizes 
\begin{equation}
    -W_\mathrm{out} = \oint_0^\tau \mu^{-1}  \dot{\gamma}^i \dot{\gamma}^j g_{ij}\big( \gamma(t) \big) + \dot{\gamma}^i A_i \big( \gamma(t) \big) \dt. \label{SM-eq:output-work}
\end{equation}
(We must confine to trajectories $\gamma(t) \in \mathcal{N} = \mathcal{M}$ in that we are only allowed to measure $g_{ij}$ and $A_i$ on $\mathcal{M}$).

\textbf{B.} Calculate the counterdiabatic term $\eta(t)$ that solves
\begin{equation}
    \frac{\partial \rhost_{\gamma^*(t)}}{ \partial t} = \mu \boldnabla \cdot (\rhost_{\gamma^*(t)} \boldnabla V_t) = \eta^i \big\{ \mu \boldnabla \cdot (\rhost_{\gamma^*(t)} \boldnabla U_i) \big\} \label{eq:SM-counterdiabatic-eq}
\end{equation}
through a projection method (c.f., \cite{zhong2024beyond}) 
\begin{equation}
    \eta^i(t) = A^{ij} \big(\gamma(t)\big) B_{j k} \big(\gamma(t)\big)  \dot{\gamma}^k(t)
\end{equation} 
for calculable tensors $A$ and $B$ [Eqs.~(S20) and (S21)].

\textbf{C.} Optimal protocols are formed as the sum of the optimal thermodynamic state trajectory and counterdiabatic terms
\begin{equation}
    \lambda^*(t) = \gamma^*(t) + \eta(t). 
\end{equation}

These three steps resemble Alg.~1 in \cite{zhong2024beyond}, but are made much more nuanced by $\boldFnc(\boldx) \neq 0$. Now we dive into detail for each of the three steps.

\subsection{Calculating $\gamma^*(t)$}

In order to calculate the optimal $\tau$-periodic curve $\gamma^*(t)$ that minimizes Eq.~\eqref{SM-eq:output-work}, we have to be able to evaluate $g_{ij}(\gamma)$ and $A_i(\gamma)$ for $\gamma \in \mathcal{M}$.

For the thermodynamic one-form, we readily calculate $A_i = A^\mathrm{qs}_i$ where, per definition [Eq.~(S7)], 
\begin{equation}
    A^\mathrm{qs}_i(\gamma) := \bigg\langle  \frac{\partial U_{\gamma}}{\partial \gamma^i} \bigg \rangle^\mathrm{st}_{\gamma} = \big\langle  U_i \big \rangle^\mathrm{st}_{\gamma}. 
\end{equation}

Following \cite{zhong2024beyond}, the thermodynamic metric $g_{ij}(\gamma)$ may be expressed in terms of a time-correlation function 
\begin{equation}
   g_{ij}
(\gamma) = \kBT \int_0^\infty \big\langle\delta  s_i\big( {\boldx}(t')\big) \delta  s_j\big( {\boldx}(0)\big) \big\rangle_\gamma^\mathrm{sym} \dt' \label{eq:g-ij-correlation-functions}
\end{equation}
where $\delta s_i(\boldx) = \partial \ln \rhost_\gamma (\boldx)  / \partial \gamma^i$ is known as the parameter score-function or the Hatano-Sasa Y-Value \cite{zulkowski2013optimal}, and $\langle \cdot \rangle^\mathrm{sym}_{\gamma}$ is a stationary-state average under the symmetrized Langevin dynamics $\dot{\boldx} = \mu \kBT \, \boldnabla \ln \rhost_\gamma(\boldx) + \sqrt{2 \mu\kBT } \, \boldzeta(t)$ \cite{hatano2001steady} \footnote{This may be seen in that $\partial_t \rhost_{\gamma(t)} = \mu \boldnabla \cdot (\rhost_{\gamma(t)} \boldnabla V_t)$ is equivalently $\partial_t \ln \rhost_{\gamma(t)} = \mu \nabla^2 V_t + \mu \boldnabla \ln \rhost_{\gamma(t)} \cdot \boldnabla V_t =: (\kBT)^{-1}  \{ \mathcal{L}^\mathrm{sym}_\gamma \}^\dagger [V_t]$ where $\{ \mathcal{L}^\mathrm{sym}_\gamma \}^\dagger$ is the adjoint operator to these symmetrized Langevin dynamics. All else follows, as $\int |\mu \boldnabla V_t|^2 \rhost_\gamma \dx = - \int \mu V_t \boldnabla \cdot ( \rhost_\gamma \mu \boldnabla V_t) \dx $ }. We note that the before-studied linear-response-approximate expression is similar in flavor $\dot{\lambda}^i \langle \partial \ln \rhost_\lambda/\partial \lambda^i  \rangle \approx \dot{\lambda}^j \int_0^\infty  \dt'  \big\langle\delta  s_i\big( {\boldx}(t')\big) \delta  s_j\big( {\boldx}(0)\big) \big\rangle_\gamma$, with the only difference between using the correlation function betwee  symmetrized Langevin dynamics to obtain this exact relationship, c.f., using the nonconservative Langevin dynamics. 

When $\boldFnc(\boldx) \neq \boldzero$, Eq.~\eqref{eq:g-ij-correlation-functions} is difficult to evaluate without knowing the stationary state $\rhost_\gamma(\boldx)$ corresponding to $U_\gamma(\boldx)$; however, it may be approximated by symmetrizing the linear response friction tensor pioneered by \cite{sivak2012thermodynamic}
\begin{gather}
    g_{ij} \approx (g^\mathrm{lr}_{ij} + g^\mathrm{lr}_{ji})/2,
\end{gather}
where
\begin{gather}
    g^\mathrm{lr}_{ij}(\gamma) = \kBT \int_0^\infty \big\langle \delta f_i\big( {\boldx}(t')\big) \delta f_j\big( {\boldx}(0)\big) \big\rangle_\gamma^\st \dt'.
\end{gather}
Here $\delta f_i=  - \frac{1}{\kBT}\big[ \frac{\partial U_{\gamma}}{\partial \gamma_i} - \big\langle \frac{\partial U_{\gamma}}{\partial \gamma_i} \big\rangle^{\st}_{{\gamma}} \big]$ is the temperature-normalized excess conjugate force, equaling $\delta f_i = \delta s_i$ when $\boldFnc = \boldzero$. This approximation may be shown to be perturbatively correct for weak nonconservative forcing to order $o(\|\boldFnc \|^2)$. Otherwise, evaluating Eq.~\eqref{eq:g-ij-correlation-functions} requires numerically estimating the stationary state spatial score function $\boldnabla \ln {\rhost_\gamma}(\boldx)$ through sophisticated modern machine learning methods \cite{rosa2024variational, boffi2024deep}. 

Instead of calculating the trajectory $\gamma^*(t)$ through a shooting method for the Euler-Lagrangian equation $2\mu^{-1} \big( {g}_{ij} \{ \ddot{\gamma}^j + \Gamma_{a b}^j \,\dot{\gamma}^a \dot{\gamma}^b \} \big)     =   \dot{\gamma}^k ( \partial_{\gamma^i} A_k - \partial_{\gamma^k} A_i )$, where $\Gamma_{a b}^i$ are the Christoffel symbols \cite{carroll2019spacetime}, it is much more practical to variationally optimize Eq.~\eqref{SM-eq:output-work} through parameterizing the curve $\gamma_\theta(t)$ (e.g., in terms of a Fourier basis $\gamma^i_\theta(t) = \sum_\ell \{ \theta^{\ell, k,\cos} \cos (2\pi k t/\tau) +  \theta^{\ell, i, \sin } \sin (2\pi k t/\tau)\} $) and calculating $W_\mathrm{out}(\theta)$ through calculating the integral Eq.~\eqref{SM-eq:output-work} directly. This avoids having to (1) calculate the Christoffel symbols and partial derivatives of $A_i$, and (2) delicately tune shooting parameters so that the trajectory $\gamma(t)$ closes in on itself at $t = \tau$ which requires both $\gamma(\tau) = \gamma(0)$ \emph{and} $\dot{\gamma}(\tau) = \dot{\gamma}(0)$.

\subsection{Calculating $\eta(t)$ from $\gamma^*(t)$}

Given an optimal protocol $\gamma^*(t)$ corresponding to $U_\gamma(\boldx) = U_\mathrm{fixed}(\boldx) + \gamma^i U_i(\boldx)$, we need to now calculate $\eta(t)$ corresponding to $V_t(\boldx) = \eta^i U_i(\boldx)$, that is to satisfy Eq.~\eqref{eq:SM-counterdiabatic-eq}. The left hand side may be expanded as 
\begin{equation}
   \frac{\partial \rhost_{\gamma^*(t)}}{ \partial t} = \dot{\gamma}^i \frac{\partial \rhost_{\gamma(t)}}{\partial \gamma^i} = -\dot{\gamma}^i \mathcal{L}_\gamma^{-1} \{\boldnabla \cdot (\rhost_{\gamma(t)} \mu \boldnabla U_i) \}
\end{equation}
wherein $ \mathcal{L}_\gamma^{-1} $ is the Drazin inverse of the Fokker-Planck operator---this comes from the identity $\mathcal{L}_\gamma [\rhost_\gamma] = 0 \rightarrow \partial _{\gamma^i} \{\mathcal{L}_\gamma [\rhost_\gamma] \} = \partial _{\gamma^i} \{\mathcal{L}_\gamma \} [\rhost_\gamma] + \mathcal{L}_\gamma [\partial _{\gamma^i}\rhost_\gamma] = \boldnabla \cdot (\rhost_\gamma \mu \boldnabla \partial_{\gamma^i} U_\gamma) + \mathcal{L}_\gamma [\partial _{\gamma^i}\rhost_\gamma]  = 0 \rightarrow \mathcal{L}_\gamma [\partial _{\gamma^i}\rhost_\gamma] =  -\boldnabla \cdot (\rhost_\gamma \mu \boldnabla U_i)$. 

Plugging this back into  Eq.~\eqref{eq:SM-counterdiabatic-eq}, we are solving for $\eta$ that satisfies 

\begin{equation}
    \eta^i \big\{ \mu \boldnabla \cdot (\rhost_{\gamma^*(t)} \boldnabla U_i) \big\} = -\dot{\gamma}^k \mathcal{L}_\gamma^{-1} \{\boldnabla \cdot (\rhost_{\gamma(t)} \mu \boldnabla U_k) \}.
\end{equation}

Finally, by taking the inner product with $-U_j(\boldx) $, (that is, $-\eta^i \int U_j \big\{ \mu \boldnabla \cdot (\rhost_{\gamma^*(t)} \boldnabla U_i) \big\} \dx = \dot{\gamma}^k  \int U_j\, \mathcal{L}_\gamma^{-1} \{\boldnabla \cdot (\rhost_{\gamma(t)} \mu \boldnabla U_k) \} \dx$) we get 
\begin{equation}
    \eta^i = A^{ij}(\gamma) B_{jk}(\gamma) \,\dot{\gamma}^k .
\end{equation}
Here we denote $A^{ij} = [A^{-1}]^{ij}$, with the tensors $A$ and $B$ defined as
\begin{equation}
    A_{ij}(\gamma) = \mu \int \rhost_\gamma(\boldx) \{ \boldnabla U_i(\boldx) \cdot \boldnabla  U_j(\boldx)\} \dx = \mu \big\langle \boldnabla U_i\cdot \boldnabla  U_j \big\rangle^\mathrm{st}_{\gamma},
\end{equation}
and 
\begin{equation}
    B_{jk} = -\int U_j \mathcal{L}_\gamma^{-1} \big[ \boldnabla \cdot (\rhost_\gamma \mu \boldnabla U_k) \big] \dx =  \frac{1}{\kBT } \bigg\{  \big\langle \delta U_j \, \delta  U_k \big\rangle^\mathrm{st}_{\gamma} + \int_0^\infty \big\langle  U_j\big(\boldx(t') \big) \, \dot{\boldx}(0) \cdot \boldnabla U_k(\boldx(0)\big) \big\rangle^\mathrm{st}_\gamma  \dt' \bigg\} 
\end{equation}
wherein we have used the verifiable relationship $\mathcal{L}_\gamma[\rhost_\gamma(\boldx) \delta U_k(\boldx)] =  \kBT \boldnabla \cdot (\rhost(\boldx) \mu \boldnabla U_k) - \boldnabla \cdot (\rhost \boldvst \cdot \boldnabla U_k)$. 

In the case where $\boldFnc = \boldzero$, the tensor $A_{ij}$ becomes the Onsager transport coefficient matrix, while the tensor $B_{jk}$ is the Fisher-information matrix. 

\subsection{Constructing the protocol $\lambda(t) = \gamma(t) + \eta(t)$}
Finally, after obtaining $\gamma^*(t)$ and $\eta(t)$, the optimal protocol may be constructed as $\lambda(t) = \gamma(t) + \eta(t)$. In essence, we have reduced solving for nonequilibrium protocols $\lambda(t)$ into procedural steps (A, B, and C) that avoid having to directly solve partial differential equations. The counterdiabatic term $\eta(t)$ is required to go beyond linear response, and has shown improvement over using $\lambda(t) = \gamma(t)$ alone \cite{zhong2024beyond}.

\section{{Zero housekeeping heat for tight confinement.}}

{Here we demonstrate that the housekeeping heat for the stationary state of an infinitely stiff and narrow potential, e.g., of the form }
\begin{equation}
    U_\kappa(\boldx) = U_0(\boldx)  + \kappa  |\boldx - \boldr_0|^2/2 \label{eq:confining-potential}
\end{equation}
and taking the limit $\kappa \rightarrow \infty$, produces zero-housekeeping heat. Given that $\boldsymbol{F}_0  = \boldFnc - \boldnabla U_0$ is suitably differentiable around $\boldx = \boldr_0$ \cite{lipschitz}, the stationary state distribution approaches a Gaussian for large values $\kappa$
\begin{equation}
    \rhost_{\kappa}(\boldx) \approx \bigg( \frac{\kappa }{2 \pi \kBT  } \bigg) ^{d/2} \exp \bigg(- \frac{\kappa | \boldx -  \boldr_0 - \kappa^{-1}\boldFzero(\boldr_0)|^2}{2 \kBT } \bigg),
\end{equation}
while the steady state velocity field approaches 
\begin{equation}
    \boldvst_\kappa(\boldx) = \mu \big\{ \boldFnc(\boldx) - \boldFnc(\boldr_0) \big\} + o(\kappa^{-1}). 
\end{equation}
The housekeeping dissipation rate $\sigma_\mathrm{hk} = \mu^{-1} \int \rhost_\kappa (\boldx )|\boldvst_\kappa (\boldx)|^2 \dx $ may be computed as a Gaussian integral
\begin{equation}
    \sigma_\mathrm{hk}(\kappa) =  \frac{\mu \kBT  \, \mathrm{tr} (  \boldnabla \boldFnc^T \boldnabla \boldFnc)}{\kappa} \bigg|_{\boldx = \boldr_0} + o(\kappa^{-2}), 
\end{equation}
where $\boldnabla \boldFnc$ is the Jacobian of $\boldFnc(\boldx)$, and $\mathrm{tr}(\cdot)$ is the matrix trace. It is readily seen that in the limit $\kappa \rightarrow \infty$, that $\rhost_\kappa \rightarrow  \delta (\boldx - \boldr_0)$, and $\sigma_\mathrm{hk}(\kappa) \rightarrow 0$.

{Note that in this $\kappa_\gamma \rightarrow \infty$ limit, the thermodynamic state approaches $\lim_{\kappa_\gamma \rightarrow \infty} \rhost_\gamma(\boldx) = \delta(\boldx -  \boldr_\gamma )$ \cite{blaber2022optimal}, and so the output work [Eq.~(14) in main text] may be written with $A^\mathrm{nc}_i$ [Eq.~(18) in main text] as}
\begin{equation}
    -W_\mathrm{out} = \oint_0^\tau \big[ \mu^{-1} |\dot{\boldr}_{\gamma(t)}|^2 - \dot{\boldr}_{\gamma(t)}\cdot \boldFnc({\boldr}_{\gamma(t)} ) \big] \dt, \label{eq:work-out-dirac-delta}
\end{equation}
{which is precisely an Euclidean Lorentz force Lagrangian action for $\boldr_{\gamma(t)}$ with vector potential $\mathbf{A}(\boldr) = -\boldFnc(\boldr)$ and charge-to-mass ratio $q/m = \mu/2$ \cite{tighttransport}.}

\section{Thermodynamic geometry terms for rotational-translational parameterization}

In this section we derive for $\gamma = (\alpha, \Sigma)$ with $\rhost_{(\alpha, \Sigma)}(\theta) = \rhost_\Sigma(\theta - \alpha)$, that
\begin{equation}
    g_{\alpha \alpha} = C,  \ \  A_\alpha = -C \Fnc,   \ \ \mathrm{and} \ \ \sigma_\mathrm{hk} =  (1 - C)\Fncsquared, \label{eq:SM-particle-on-a-ring-thermo-geo} 
\end{equation}
where
\begin{equation}
    C(\Sigma) =  1 - \frac{4\pi^2}{ \int_0^{2\pi}  \{ \rhost_{\Sigma}(\theta) \} ^{-1}   \dvarphi}  \in [0, 1] \label{eq:SM-concentration-constant}
\end{equation}
is a concentration constant for $\rhost_\Sigma(\theta)$. For notational simplicity, we consider $\Sigma = \Sigma_0$ constant, and we notate here $\rho^\st_{\alpha(t)}(\theta) := \rho^\st_{\Sigma_0}\big( \theta - \alpha(t) \big)$, and set $\mu = 1$ to be restored later.

The main observation is that ${V}_t(\gamma)$ may be solved for quasi-analytically
\begin{align}
    \partial_\theta (\rhost_{\alpha} \partial_\theta {V}_t) = \partial_t \rhost_{\alpha(t)} =  -\dot{\alpha} \,\partial_\theta \rhost_{\alpha(t)} ,
\end{align}
the second equality from $\rho^\st_{\alpha(t)}(\theta) = \rho^\st_{0}\big( \theta - \alpha(t)\big)$. Together this yields
\begin{align}
    \partial_\theta \{ \rhost_\alpha (\partial_\theta {V}_t + \dot{\alpha} ) \} &= 0  \\
    \Longrightarrow \quad\quad\quad \rhost_\alpha (\partial_\theta {V}_t + \dot{\alpha} ) &= c \quad (\mathrm{constant}) \\ 
    \Longrightarrow \quad\quad\quad \partial_\theta {V}_t &= -\dot{\alpha} + c \big(\rhost_\alpha\big)^{-1}.
\end{align}
In order to solve for $c$, we apply $\int_0^{2\pi} \partial_\theta \,b(\theta) \dvarphi = 0$ for all $2\pi$-periodic scalar functions $b(\theta)$ to the above
\begin{align}
   0 &= -2\pi \dot\alpha + c \int_0^{2\pi} \big(\rhost_\alpha\big)^{-1} \dvarphi  \\
    \Longrightarrow \quad\quad \quad c &= \frac{2\pi \dot{\alpha}}{\int_0^{2\pi} \big(\rhost_\alpha\big)^{-1} \dvarphi} \\ 
    \Longrightarrow \quad\quad \quad \partial_\theta {V}_t &= -\dot{\alpha} \bigg\{1 - \frac{2\pi \big( \rhost_\alpha \big)^{-1}}{\int_0^{2\pi} \big[\rhost_\alpha\big(\theta')]^{-1} \dvarphi'}  \bigg\}
\end{align}
Plugging this into Eq., we can now evaluate 
\begin{align}
    \dot{\alpha}^2 g_{\alpha \alpha} &= \int_0^{2\pi} \rhost_\alpha (\partial_\theta {V}_t)^2 \dvarphi \\
    &= \dot{\alpha}^2 \int_0^{2\pi} \rhost_\alpha \bigg\{1 - \frac{2\pi \big( \rhost_\alpha \big)^{-1}}{\int_0^{2\pi} \big[\rhost_\alpha\big(\theta')]^{-1} \dvarphi'}  \bigg\}^2 \dvarphi
\end{align}
Expanding the square and integrating yields ultimately
\begin{align}
    g_{\alpha\alpha} &= 1 -  \frac{4\pi^2}{\int_0^{2\pi} \big[\rhost_\alpha\big(\theta')]^{-1} \dvarphi'}, 
\end{align}
which is the concentration constant $C(\Sigma_0)$ as defined in Eq.~\eqref{eq:SM-concentration-constant}. 

We can use a similar trick to solve for $\vst_\alpha(\theta) = \Fnc - \partial_\theta(U_\alpha + \ln \rhost_\alpha)$ through 
\begin{align}
     \partial_\theta (\rhost_\alpha [\partial_\theta(U_\alpha + \ln \rhost_\alpha) - \Fnc]) &= 0 \\ 
    \Longrightarrow \quad\quad \rhost_\alpha [\partial_\theta(U_\alpha + \ln \rhost_\alpha) - \Fnc] &= \tilde{c} \quad \mathrm{(constant)} \\ 
    \Longrightarrow \quad\quad\partial_\theta (U_\alpha + \ln \rhost_\alpha) &= \tilde{c} \big( \rhost_\alpha\big)^{-1} + \Fnc 
\end{align}
We can solve again for the constant $\tilde{c}$ with the same trick, obtaining
\begin{equation}
    \tilde{c} = -\frac{2\pi \Fnc}{\int_0^{2\pi} \big(\rhost_\alpha\big)^{-1} \dvarphi},
\end{equation}
which may be plugged back in to get
\begin{equation}
  \partial_\theta (U_\alpha + \ln \rhost_\alpha) = \Fnc \bigg(1 - \frac{2\pi \big(\rhost_\alpha\big)^{-1} }{\int_0^{2\pi} \big[\rhost_\alpha(\theta')\big]^{-1} \dvarphi' }  \bigg)    
\end{equation}
yielding the stationary state velocity field
\begin{equation}
    \vst_\Sigma(\theta) = \Fnc \bigg( \frac{2\pi \big[\rhost_\alpha(\theta) \big]^{-1}}{\int_0^{2\pi} \big[\rhost_\alpha\big(\theta')]^{-1} \dvarphi'} \bigg).
\end{equation}

This leads to the housekeeping rate 
\begin{align}
    \sigma_\mathrm{hk} &= \int_0^{2\pi} \rhost_\alpha (\vst_\alpha)^2 \dvarphi  \\ 
    &= \Fncsquared \frac{4 \pi^2}{ \int_0^{2\pi} \big[\rhost_\alpha\big(\theta')]^{-1} \dvarphi'}
\end{align}
and finally, it follows that for $\dot{\alpha} A_\alpha^\mathrm{nc} = \int  \rhost_\alpha \Fnc \partial_\theta V_t \dvarphi $ that
\begin{equation}
  A_\alpha^\mathrm{nc} = \Fnc \bigg(1 -  \frac{4\pi^2}{\int_0^{2\pi} \big[\rhost_\alpha\big(\theta')]^{-1} \dvarphi'} \bigg). 
\end{equation}

Finally, restoring the $\mu$ via dimensional analysis, yields Eq.~\eqref{eq:SM-particle-on-a-ring-thermo-geo}.

\section{Deriving $g_{\alpha \Sigma} = 0$ implies $A^\mathrm{nc}_\Sigma = 0$.}

In this section, we show  $g_{\alpha \Sigma} = 0$ implies $A^\mathrm{nc}_\Sigma = 0$. Suppose $\Upsilon_\alpha(\theta)$ and $\Upsilon_\Sigma(\theta)$ are scalar fields satisfying
\begin{equation}
   \frac{\partial \rhost_\gamma(\theta) }{\partial \Sigma} = \partial_\theta \{ \rhost_\gamma(\theta) \, \mu \, \partial_\gamma \Upsilon_\Sigma(\theta) \, \}
\end{equation}
and
\begin{align}
   \frac{\partial \rhost_\gamma(\theta) }{\partial \alpha}  = \partial_\theta \{ \rhost_\gamma(\theta) \, \mu \, \partial_\theta \Upsilon_\alpha(\theta) \, \}.  = -\partial_\theta \rhost_\gamma(\theta) \label{eq:rhost-varphi-relationship}
\end{align}
because $\alpha$ is defined so that $\rhost_{(\alpha, \Sigma)}(\theta) = \rhost_{(0, \Sigma)}(\theta - \alpha)$. Note the relationship $V_t(\theta) = \dot{\alpha} \Upsilon_\alpha(\theta) + \dot{\Sigma} \Upsilon_\Sigma(\theta)$. 

We can manipulate through integration by parts in $\theta$ and substitutions of Eq.~\eqref{eq:rhost-varphi-relationship} to get
\begin{align}
    g_{\alpha \Sigma}(\gamma) &:=  \int \rhost_\gamma  (\mu \partial_\theta \Upsilon_\Sigma )  ( \mu \partial_\theta \Upsilon_\Sigma) \dvarphi  \\
    &= - \int \mu \Upsilon_\Sigma \partial_\theta \{ \rhost_\gamma \mu ( \partial_\theta \Upsilon_\alpha ) \} \dvarphi  \\
     &= \int \mu \Upsilon_\Sigma \partial_\theta \rhost_\gamma \dvarphi = - \int \rhost_\gamma \partial_\theta( \mu \Upsilon_\Sigma)  \dvarphi  \\
     &= -A^\mathrm{nc}_\Sigma(\gamma) / \Fnc, 
\end{align}
referring back to the definition for $A_\mathrm{nc}$ in the main text [Eq.~(19)], and noting that $\Fnc$ is a scalar constant in this $d=1$ case. 

Thus, if $\Sigma$ is chosen (e.g., with the Gram–Schmidt procedure) so that $g_{\alpha \Sigma} = 0$ everywhere, then $A^\mathrm{nc}_\Sigma = 0$ everywhere as well. 

\bibliography{main}